\documentstyle[12pt]{article}

\begin{document}

%%-move to normal A4-%%
\hoffset = -1truecm
\voffset = -2truecm
\baselineskip = 10 mm
\title{\bf Space-time properties of the higher twist amplitudes}
\author{
{\bf Wei Zhu$^{1,2}$}, {\bf Hongwei Xiong$^2$} and
{\bf Jianhong Ruan$^{2,3}$}\\
\normalsize $^1$CCAST(World Laboratory), P.O.Box 8730,
Beijing 100080, P.R. China \\
\normalsize $^2$Department of Physics, East China Normal University,
Shanghai 200062, P.R. China \\
\normalsize $^3$Institute of High Energy Physics, Chinese 
Academy of Sciences, Beijing 100039, P.R. China
}
\date{}

\newpage

\maketitle

\vskip 3truecm

\begin{abstract}
	
	A consistent and intuitive description of the twist-4 corrections 
to the hadron structure functions is presented in a QCD-improved parton 
model using time-ordered perturbative theory, where the collinear 
singularities are naturally eliminated. We identify the special propagators 
with the backward propagators of partons in time order. \\ \\ 


\end{abstract}

\newpage
\begin{center}
\section{Introduction}
\end{center}

	The twist-4 or inverse-power $1/Q^2$ corrections to scaling 
in deep inelastic scattering (DIS) processes are an important subject 
for a precision test of QCD and the analyses of deep inelastic 
scattering (DIS) data [1-4].  Jaffe and Soldate (JS) used OPE method 
to relate the inverse-power corrections to the target matrix elements of  
local operators [2]. On the other hand, the QCD-improved parton model can 
provide a simple and intuitive description of scaling phenomena as a  
compensate method for the OPE technique. In this aspect, 
Jaffe [4] recasted the results of work [2] in parton model. Ellis, Furmanski, 
Petronzio [3] (EFP) and Qiu [4] try to construct a complete
QCD-improved parton model including the inverse-power short-distance 
corrections based on the manipulation of stand Feynman diagrams. Where 
EFP's diagram-approach needs a special complicate procedure to 
pick up the short-distance contributions. Qiu advanced 
the above technique and proposed that a set of special propagators can
isolate the short-distance effect in Feynman diagrams [4].
However, the dynamic origin of the special propagators and collinear 
safeness of the theory are not yet clear. Obviously, we still lack a 
consistent QCD-improved parton model to analyze the inverse-power 
corrections in DIS processes. 

	In this work, we try to realize the twist-4 short-distance
corrections to the hadron structure functions using time ordered perturbative
theory (TOPT). We find that the collinear singularity can be naturally 
eliminated if we generalize TOPT to the collinear limit. In this framework, 
the twist-4 short-distance corrections to the hadron structure functions 
will present an intuitive space-time picture. In particular, we identify 
Qiu's special propagators [4] with the backward propagators of partons in 
time order. Our new method provides a consistent way to study higher-twist 
processes in a QCD-improved parton model.

	The outline of the paper is as follows. In Sec. 2, we shall 
generalize TOPT to the collinear limit. We find that the collinear and 
infrared singularities in amplitude will transfer to the interaction-time 
from the momentum-space in TOPT and they can be natural eliminated. Using 
this TOPT-framework, we shall give a set of calculating rules, which pick 
up the inverse-power short-distance corrections to the hadron 
structure functions in Sec. 3. We shall discuss the factorization scheme
in Sec. 4. A summary is given in the last section. 

\begin{center}
\section{Collinear safeness}
\end{center}
	
	As we know, TOPT is a natural way for studying space-time properties 
of the twist-2 DGLAP evolution equation [5]. In principle, a Feynman 
diagram can be decomposed to all possible time-ordered graphs in TOPT. 
However, in the leading-twist processes many TOPT amplitudes, which contain 
the backward-propagating partons, are suppressed in the infinite momentum 
frame (see Fig. 1). The reason is straightforward. The cross section of DIS 
is generally a function of the longitudinal momenta, $x_iP$ and transverse 
momenta, $l_\perp^2$, of partons. The terms with $x_iP$ are canceled in 
Fig. 1a due to the conservation of longitudinal momenta at every vertex 
and the final results depend only on $l_\perp^{-2}$ and give 
$\log Q^2$-corrections. Any backward-propagating parton in the splitting 
vertex will violate the balance among $x_iP$ in the energy terms and lead 
to $d l^2_\perp/P^2\rightarrow 0$. Thus, the backward-propagating partons 
are suppressed and the scattering amplitudes only contain the 
forward-propagating partons in the leading twist processes.  In this work 
we find that the twist-4 amplitudes also expose the simple TOPT-form but 
the contributions of backward moving partons are no longer neglected. 

	The higher twist corrections to structure functions are separated 
in the collinear expansion, where all transverse momenta of partons  
are taken as zero [4]. Therefore, we first generalize the TOPT-amplitude 
to the collinear limit. As an example, we consider the process 
$p_1+p_2\rightarrow p_1'+p_2'$, in which the parton momenta are parametrised 
as 

$$p_1:(p_{10}, l_\perp, p_{1z}), \hspace{0.15cm} p_2:(p_{20},-l_\perp, p_{2z}), 
\hspace{0.15cm} p_1':(p_{10}', l_\perp, p_{1z}'), \hspace{0.15cm}
P_2':(p_{10}',-l_\perp, p_{2z}'), \hspace{0.15cm} 
k:(k_0, 0_\perp, k_z) \eqno(1) $$ 
where the transverse momenta $l_\perp$ will be set to zero. We emphasize 
that our following results are independent of the parameters (1). 
We choose a coordinate frame such that 
	
$$p^\mu=p\overline{n}^\mu, $$
$$q^\mu=-x_Bp\overline{n}^\mu+\frac{Q^2}{2x_Bp}n^\mu. \eqno(2)$$

The contributions of the Feynman propagator of fermion (quark or antiquark)
with the momentum $k$ to the amplitude are 

$$\int d^4kS(k)=\int \frac{dk_0d^3\vec{k}}{(2\pi)^4}\frac{\gamma\cdot k} 
{k_0^2-\omega^2+i\epsilon}(2\pi)^4\delta^4(p_1+p_2-k)(2\pi)^4\delta^4
(k-p_1'-p_2')$$
$$=\frac{i}{2\omega}\{\int^\infty_{-\infty}dt'e^{-it'(\omega-p_{10}'
-p_{20}')}\int^{t'}_{-\infty}dte^{-it(p_{10}+p_{20}-\omega)}
[-i\gamma\cdot k\delta(k_0-\omega)]+$$
$$\int^\infty_{-\infty}dte^{-it(p_{10}
+p_{20}+\omega)}\int^{t}_{-\infty}dt'e^{it'(p_{10}'+p_{20}'+\omega)}
[i\gamma\cdot k\delta(k_0+\omega)]\}$$
$$(2\pi)^3\delta^3(\vec{p}_1+\vec{p}_2-\vec{p}_1'-\vec{p}_2')$$
$$\equiv [S_F(k)+S_B(k)](2\pi)^4\delta^4(p_1+p_2-p_1'-p_2'), \eqno(3)$$
where $\omega=\sqrt{k_\perp^2+(k_z)^2}$. We use light-cone coordinates, in
which we define   

$$\overline{n}^{\mu}\equiv (\overline{n}^0, \overline{n}_\perp,
\overline{n}^3)=\frac{1}{\sqrt{2}}(1, 0_\perp,1);\hspace{0.3cm}
n^{\mu}=\frac{1}{\sqrt{2}}(1, 0_\perp, -1),\eqno(4)$$
or equivalently 

$$\overline{n}^{\mu}\equiv(\overline{n}^+,
\overline{n}^-,\overline{n}_\perp)
=(1, 0, 0_\perp);\hspace{0.3cm} n^{\mu}=(0,1, 0_\perp). \eqno(5)$$
Thus, $\overline{n}^2=0, n^2=0$, and $\overline{n}\cdot n=1$; 
$x=k\cdot n/P\cdot n$ is the light-cone fraction. In the collinear limit, 
we take $l_\perp\rightarrow 0$. In consequence, we have

$$\lim_{l_\perp\rightarrow 0}S_F(k)(2\pi)^4\delta^4(p_1+p_2-p_1'-p_2')$$
$$=\lim_{p_{10}+p_{20}\rightarrow \omega}\frac{\gamma\cdot k \delta(k_0-\omega)}{2\omega}\int^\infty_{-\infty}
dt'e^{-it'(\omega-p'_{10}-p'_{20})}$$
$$(\frac{e^{-it'(p_{10}+p_{20}-\omega)}}{p_{10}+p_{20}-\omega}-
\frac{e^{-iT(p_{10}+p_{20}-\omega)}}{p_{10}+p_{20}-\omega})
(2\pi)^3\delta^3(\vec{p}_1+\vec{p}_2-\vec{p}_1'-\vec{p}_2')$$
$$=\frac{\sqrt{2}\omega\gamma \cdot \overline{n}}{2\omega}\int^\infty_{-\infty}
dt'e^{-t'(\omega-p_{10}'-p_{20}')}(t'-T)(2\pi)^3\delta^3(\vec{p}_1
+\vec{p}_2-\vec{p}_1'-\vec{p}_2')$$
$$=\frac{\sqrt{2}\alpha \gamma\cdot \overline{n}}{2\omega}(2\pi)^4\delta^4(p_1+p_2
-p_1'-p_2'). \eqno(6)$$

	Usually, the lower limit T of the integral, which is relating to 
the interaction time, is taken as $-\infty$. However, T is really a 
finite quantity in physics and we denote $t'-T\equiv \alpha/\omega$, where
$\alpha$ is a dimensionless quantity. We emphasis that the term  
$-e^{-iT(p_{10}+p_{20}-\omega)}/(p_{10}+p_{20}-\omega)$ in Eq. (6), 
which vanishes in the noncollinear case, is important for the collinear  
safeness. 

	The second term in Eq. (3) is
	
$$\lim_{l_\perp\rightarrow 0}S_B(k)(2\pi)^4\delta^4(p_1+p_2-p_1'-p_2')$$
$$=\lim_{p_{10}'+p_{20}'\rightarrow \omega}-\frac{\gamma\cdot k \delta(k_0+\omega)}
{2\omega}\int^\infty_{-\infty}
dte^{-it(\omega+p_{10}+p_{20})}
\frac{e^{it(p_{10}'+p_{20}'+\omega)}}{p_{10}'+p_{20}'+\omega}
(2\pi)^3\delta^3(\vec{p}_1+\vec{p}_2-\vec{p}_1'-\vec{p}_2')$$
$$=-\frac{\gamma\cdot k\delta(k_0+\omega)}{4\omega^2}(2\pi)^4\delta^4(p_1+p_2
-p_1'-p_2')$$
$$=\frac{\gamma\cdot n}{2k\cdot n}(2\pi)^4\delta^4(p_1+p_2
-p_1'-p_2'). \eqno(7)$$
	
	The negative frequency solution $k_0=-\omega$ in $S_B(k)$ can
be understand as a backward-moving parton in time order since 
$k\delta(k_0+\omega)=-(\omega,0_\perp,-k_z)\equiv \hat{k}$, $\hat{k}$ is
on-shell momentum in TOPT. We will refer   
to $S_F(k)$ and $S_B(k)$ as the forward and backward propagators,
respectively. We can generalize $\hat{k}$ to the off-shell momentum k  
in (6-7) since $n\cdot n=0$, where the transverse momentum of parton 
is vanished. Obviously, the backward propagator
of quark is identical with the special propagator [4] or the instantaneous
part of quark propagator in the non-collinear case [6]. 
Qiu has pointed out that the special propagator offers no space separation
along the light-cone coordinate of two interaction points connected by 
the propagator. One can straightforwardly understand the above mentioned
property as follows: the propagation of a backward parton "seems" 
to be instantaneous in kinematics when $ P\rightarrow \infty$. It is 
interesting that Fig. 2b shows that the virtual photon creates 
a $q\overline {q}$ pair in the infinite-momentum frame.
	
	In general, the massless partons with the parallel        
momenta can go on-mass-shell simultaneously in the collinear case 
and collinear singularities may arise. Therefore, one might think that  
divergence could arise in the forward propagator $S_F(k)$ 
if $k^2=0$ (we call it as the collinear propagator). 
Fortunately, Eq. (6) shows that the collinear singularity is absent due to 
the compensative contributions from the factor  
$-e^{-iT(p_{10}+p_{20}-\omega)}/(p_{10}+p_{20}-\omega)$ and $\alpha/\omega$
is a finite quantity in Eq. (6). Furthermore, due to helicity conservation,
the contributions of the numerator containing the collinear forward 
propagator to the twist-4 structure functions vanish. For example, the 
emission of an on-shell massless collinear gluon along the quark line 
is forbidden in the collinear approximation. Therefore, 
one can neglect the collinear propagator $S_F(k)$ in the twist-4 component of 
structure function provided $\alpha/\omega$ is finite.
Thus, we show a new way to eliminate collinear divergence: the 
collinear singularities will be transferred from the momentum-space to the 
interaction-time in TOPT and the later is really finite in physics. 

	Now let us discuss the gluon propagator $G^{\alpha\beta}(k)$.
Similar to Eqs. (3-7), we have 

$$\lim_{l_\perp\rightarrow 0}\int d^4kG^{\alpha\beta}(k)$$ 
$$=\frac{2\alpha \Gamma^{\alpha\beta}(k)\delta(k_0-\omega)}{(k\cdot n)^2}
(2\pi)^4\delta^4(p_1+p_2-p_1'-p_2')
-\frac{\Gamma^{\alpha\beta}(k)\delta(k_0+\omega)}
{(k\cdot n)^2}(2\pi)^4\delta^4(p_1+p_2-p_1'-p_2')$$
$$\equiv [G_F^{\alpha\beta}(k)+G_B^{\alpha\beta}(k)](2\pi)^4\delta
^4(p_1+p_2-p_1'-p_2'), \eqno(8)$$
where $\Gamma^{\alpha\beta}$ is the gauge-dependent polarization sum.
In the light-cone gauge and limit $l_\perp\rightarrow 0$,

$$\Gamma^{\alpha\beta}(k)\delta(k_0-\omega)
=-g^{\alpha\beta}+\frac{k^\alpha n^\beta+k^\beta n^\alpha}{2\omega}
\delta(k_0-\omega)$$
$$=\delta^{ij}_\perp, \eqno(9)$$
in $G_F^{\alpha\beta}(k)$

and

$$-\Gamma^{\alpha\beta}(k)\delta(k_0+\omega)
=g^{\alpha\beta}-\frac{k^\alpha n^\beta+k^\beta n^\alpha}{2\omega}
\delta(k_0+\omega)$$
$$=g^{\alpha\beta}+n^\alpha n^\beta, \eqno(10)$$
in $G_B^{\alpha\beta}(k)$, respectively. Where we use $\omega=xp$. Therefore,

$$G_F^{\alpha\beta}(k)
=\frac{2\alpha \delta^{ij}_\perp}{(k\cdot n)^2}, \eqno(11)$$

and       

$$G_B^{\alpha\beta}(k)
=\frac{g^{\alpha\beta}}{(k\cdot n)^2}+\frac{n^\alpha n^\beta}{(k\cdot n)^2}, 
\eqno(12)$$
where $\delta^{ij}_\perp$ and $g^{\alpha\beta}$ collect the contributions
of the terms with $l^2_\perp$ in the numerator. Thus, in the collinear limit, 
we need only consider the contributions of 

$$G_B^{\alpha\beta}(k)
=\frac{n^\alpha n^\beta}{(k\cdot n)^2} \eqno(13)$$
in calculations of the twist-4 structure functions.
Equation (13) is the special propagator $G^{\alpha\beta}_s(k)$ [4] or the 
instantaneous propagator for gluon in the noncollinear case [6].

	From the above discussions, we can conclude that the 
contributions of the vertices (Figs. 3a and 3b) vanish due to collinear 
safeness, where F and B label the forward- and backward-moving partons, 
respectively. Furthermore, the vertices in Figs. 3c and 3d also are forbidden due to 
$\gamma\cdot n\gamma\cdot n=0$ and $n\cdot n=0$.

	The amplitude with twist-4 corrections in the collinear 
expansion can contain the noncollinear propagators with off-shell momenta
although parton transverse momenta are neglected. 
Any noncollinear propagators can be decomposed to a non-vanishing 
forward and backward components in TOPT, in which $S_B(k)$ and
$G_B^{\alpha\beta}(k)$ have the same form as (7) and (13), respectively. On 
the other hand, we can use stand TOPT to calculate $S_F(k)$ and $G_F(k)$
and get

	$$S_F(k)=\frac{\gamma\cdot \overline{n}}{2k\cdot \overline{n}}, 
\eqno(14)$$

and

	$$G_F^{\alpha\beta}(k)
=\frac{\delta^{ij}_\perp}{k\cdot \overline{n}(k\cdot n-k\cdot\overline{n})}, 
\eqno(15)$$
where k are off-shell momenta. From the discussions about the propagators, 
we can find that the backward-propagating gluon is longitudinally polarized, 
while the forward-propagating gluon is transversely polarized in the 
collinear approximation. A parton will inverse (or will keep) 
its moving direction after absolving a forward (or backward) gluon due to 
helicity conservation (Fig. 4). It means that the vertices of Figs. 3e-g  
are forbidden. Thus, a backward-moving gluon only couples with two 
forward-moving partons, while a forward-moving gluon couples with a pair
of quark and antiquark moving oppositely.

	The selecting rules (Fig. 4) simplify the calculations of the 
twist-4 corrections in the collinear expansion. In fact, we have the 
following TOPT-procedure for calculating the high-twist contributions  
in the collinear expansion: 
(1) to list all possible time-ordered diagrams of a photon-parton forward 
scattering amplitudes; (2) to label B or F on the parton lines according 
to Fig. 4, and note that all loop momenta are forward while the cut 
propagator is denoted by F or B for gluon or fermion, respectively, since 
the polarization sums of initial and final gluons have the same form 
$\delta^{ij}_\perp$; (3) the contributions of the backward and noncollinear 
forward propagators are calculated by 
$S_B(k)$, $G_B^{\alpha\beta}(k)$, $S_F(k)$ and $G_F^{\alpha\beta}(k)$, respectively. 
One can find that every cut-diagram only corresponds to a non-vanished 
TOPT-graph due to the coupling of partons in space-time (Fig. 4).  In 
opposition to our TOPT-method, the covariant perturbative methods [3,4] 
do not tell whether forward propagator or backward propagator is more 
important, since they mix together all possible time ordering in Feynman 
diagram. Obviously, the gauge invariance is preserved if we include $all$ 
TOPT-diagrams of a physical process at given order of coupling constant.

\begin{center}
\section{Power corrections to structure functions}
\end{center}

	One can simply reproduce all results of Ref. [4] using our 
TOPT-programs in Sec. 2. In deep inelastic scattering of lepton-hadron, 
the hadronic tensor $W^{\mu\nu}(p,q)$ can decompose to

$$W^{\mu\nu}(p,q)=e^{\mu\nu}_L(x_B,Q^2)+e^{\mu\nu}_T(x_B,Q^2), \eqno(16)$$
where $e^{\mu\nu}_L$ and $e^{\mu\nu}_T$ are the longitudinal and transverse 
tensors, respectively.  Using the definition (4), 

$$e^{\mu\nu}_L=\frac{1}{2}\left (g^{\mu\nu}+\frac{(x_Bp)^2}{Q^2}
\overline{n}^\mu\overline{n}^\nu+\frac{Q^2}{(2x_Bp)^2}n^\mu n^\nu
-\frac{1}{2}(\overline{n}^\mu\overline{n}^\nu)\right),\eqno(17)$$

$$e^{\mu\nu}_T=\frac{1}{2}\left (\overline{n}^\mu n^\nu+
n^\mu \overline{n}^\nu-g^{\mu\nu}\right)=\frac{1}{2}d^{\mu\nu}. \eqno(18)$$
The standard structure functions are defined as

$$2F_1(x_B,Q^2)=F_T(x_B,Q^2),$$

$$F_2(x_B,Q^2)=x_B[F_T(x_B,Q^2)+F_L(x_B,Q^2)]. \eqno(19) $$

The inverse-power corrections of the twist-4 QCD processes to the structure 
functions $F_1$ and $F_2$, or the hadronic tensor $W^{\mu\nu}$ are from 
Fig. 5 (for quark-gluon-to-quark-gluon) and Fig. 6 
(for two-quark-to-two-quark), respectively. In the concrete, 
$W^{\mu\nu}_4\sim\int\sigma^{\mu\nu}T$, where $\sigma^{\mu\nu}$ are the 
cross section of the processes in Figs. 5 and 6, T is the contributions 
from the target part [4,7]. Using our calculating program in TOPT, we can 
easy write $\sigma^{\mu\nu}$ in Fig. 5 as

$$\sigma^{\mu\nu}_{qga}(x,x_1,x_2)$$
$$=\frac {d^{\alpha\beta}}{16\pi}Tr[\gamma^\alpha\frac{\gamma\cdot n}
{2k\cdot n}\gamma^\mu\frac{Q^2}{2x_Bp}\gamma\cdot n\gamma^\nu
\frac{\gamma\cdot n}{2k\cdot n}\gamma^\beta p\gamma\cdot \overline{n}]
2\pi\delta(x-x_B)\frac{x_B}{Q^2}$$
$$=2n^\mu n^\nu\delta(x-x_B)(\frac{1}{2xp})^2, \eqno(20)$$

$$\sigma^{\mu\nu}_{qgb}(x,x_1,x_2)$$
$$=\frac {d^{\alpha\beta}}{16\pi}Tr[\gamma^\mu\frac{x_Bp}{Q^2}
\gamma\cdot \overline{n}\gamma^\alpha\frac{Q^2}{2x_Bp}\gamma\cdot n
\gamma^\beta\frac{x_Bp}{Q^2}\gamma\cdot\overline{n}\gamma^\nu p\gamma
\cdot \overline{n}]
2\pi\delta(x-x_B)\frac{x_B}{Q^2}$$
$$=2\overline{n}^\mu\overline{n}^\nu\delta(x-x_B)(\frac{px}{Q^2})^2, 
\eqno(21)$$

$$\sigma^{\mu\nu}_{qgc}(x,x_1,x_2)$$
$$=\frac {d^{\alpha\beta}}{16\pi}Tr[\gamma^\mu\frac{Q^2}{2x_Bp}
\gamma\cdot n\gamma^\alpha\frac{x_Bp}{Q^2}\gamma\cdot \overline{n}
\gamma^\beta\frac{\gamma\cdot n}{2(x_1-x_B)xp}\gamma^\nu p\gamma\cdot\overline{n}]
2\pi\delta(x_2-x_B)\frac{x_B}{Q^2}$$
$$=e_T^{\mu\nu}\delta(x_2-x_B)\frac{x_2}{(x_1-x_2)Q^2}, \eqno(22)$$

$$\sigma^{\mu\nu}_{qge}(x,x_1,x_2)$$
$$=\frac {d^{\alpha\beta}}{16\pi}Tr[\gamma^\mu\frac{x_Bp}{Q^2}
\gamma\cdot \overline{n}\gamma^\alpha\frac{Q^2}{2x_Bp}\gamma\cdot n
\gamma^\nu\frac{\gamma\cdot n}{2xp}\gamma^\beta p\gamma\cdot\overline{n}]
2\pi\delta(x-x_B)\frac{x_B}{Q^2}$$
$$=\overline{n}^\mu n^\nu\delta(x-x_B)\frac{1}{Q^2}. \eqno(23)$$
Note that the contributions of Figs.5g-5j are vanished since in Figs. 5g 
and 5h the propagators F$_1$ contributes 
$\sim \gamma\cdot \overline{n}/p$ and leads to $\sigma_T^{\mu\nu}/p^2
\rightarrow 0$ if $p\rightarrow \infty$; Figs. 5i and 5j include
$Tr[\gamma^\mu\gamma\cdot n\gamma^\alpha\gamma\cdot\overline{n}
\gamma^\nu\gamma\cdot n\gamma^\beta\gamma\cdot\overline{n}]=0$.

Similar, the contributions of four-quark process to the leading
inverse power corrections in Fig. 6 are

$$\sigma^{\mu\nu}_{qqa}(x,x_1,x_2)$$
$$=\frac {d^{\alpha\beta}}{16\pi}Tr[\gamma^\mu\frac{\gamma\cdot n}
{2(x_2-x_B)p}\gamma^\alpha p\gamma\cdot \overline{n}
\gamma^\beta\frac{\gamma\cdot n}{2(x_1-x_B)p}\gamma^\nu p\gamma\cdot
\overline{n}]
2\pi\delta(x-x_B)\frac{x_B}{Q^2}$$
$$=e_T^{\mu\nu}\delta(x-x_B)\frac{x}{(x_1-x)(x_2-x)Q^2}, \eqno(24)$$

$$\sigma^{\mu\nu}_{qqb}(x,x_1,x_2)$$
$$=\frac {d^{\alpha\beta}}{16\pi}Tr[\gamma^\mu\frac{\gamma\cdot n}
{2(x_2-x_B)p}\gamma^\alpha p\gamma\cdot \overline{n}
\gamma^\beta\frac{Q^2}{2(x_2-x_B)p}\gamma^\alpha p\gamma\cdot
\overline{n}]\frac{2x_1}{(x-x_1)Q^2}
2\pi\delta(x_1-x_B)\frac{x_B}{Q^2}$$
$$=e_T^{\mu\nu}\delta(x_1-x_B)\frac{x_1}{(x-x_1)(x_2-x_1)Q^2}. \eqno(25)$$
Thus, we reproduce the results of Refs. [3,4] using TOPT. They are

$$\sigma^{\mu\nu}_{qg}(x,x_1,x_2)$$
$$=\frac{8}{Q^2}(e_T^{\mu\nu}+e_L^{\mu\nu})\delta(x-x_B)-\frac{2x_B}{Q^2}
e_T^{\mu\nu}\left [\frac{\delta(x_2-x_B)-\delta(x_1-x_B)}{x_2-x_1}\right ],
\eqno(26)$$

$$\sigma^{\mu\nu}_{qq}(x,x_1,x_2)$$
$$=\frac{x_B}{Q^2}e_T^{\mu\nu}\left [\frac{\delta(x_2-x_B)-\delta(x_1-x_B)}
{(x_2-x)(x_2-x_1)}-\frac{\delta(x-x_B)-\delta(x_1-x_B)}{(x_2-x)(x-x_1)}\right.$$
$$\left.+\frac{\delta(y_2-x_B)-\delta(y_1-x_B)}
{(y_2-x)(y_2-y_1)}-\frac{\delta(y-x_B)-\delta(y_1-x_B)}{(y_2-x)(x-y_1)}
\right ], \eqno(27)$$
where $y_1=x_1-x$ and $y_2=x-x_2$.

	In the above mentioned calculations we considered the processes of
two-parton-to-two-parton. We also should pick up the contributions from
all cut diagrams of the interference process of one-parton-to-one-parton
and three-parton-to-one-parton. For example,
we need include contributions of Fig. 7, which are corresponding to Fig. 5c. 
One can find that the contributions of Figs. 7a
and 7b are the same as that of Fig. 5c. Therefore, we should multiply
the above mentioned results (26) and (27) by a factor 3 for including the
contributions of the interference processes.

\begin{center}
\section{Factorization scheme in TOPT}
\end{center}

	A relating subject with the above mentioned statements of 
power corrections to structure functions is the factorization scheme in 
TOPT. As we know that factorization theorem allows us to separate the calculable 
short-distance effects from incalculable long-distance part.
A complete derivation of the factorization theorem in DIS 
has been given by many authors [8]. The discussion of this section 
intends to provide an intuitive understanding of the existing 
factorization scheme.

	 A simplest amplitude of DIS process contains a pair of 
(left and right) loop Feynman propagators and it can be written as
	
$$M^{\mu\nu}(p,q)=Tr[\gamma\cdot pTDHNH^*D^*T^*], \eqno(28)$$
where $D$ and $D^*$ are Feynman propagators of quarks connecting with the 
target-part $T$ $(T^*)$ and hard-part $H$ $(H^*)$, respectively. 
According to the view of parton model, in a factorization scheme,
all virtual (off-shell) partonic lines joining $T$ and $H$ 
can be broken to the free (on-shell) partonic lines. 
In this case, we can re-stipulate those on-shell partons as incoming and 
outgoing partons in the target- and hard-parts, respectively.

	Any Feynman propagator is decomposed to the forward $F$ and backward 
$B$ propagators and the propagating partons are evaluated on-shell in TOPT. 
The diagram including two forward propagators in Fig. 8a gives the leading 
twist-2 contribution and we can separate the calculable 
short-distance effects from incalculable target part, since two 
on-shell-forward 
propagators can be broken to incoming and outgoing partons in the 
target and hard parts in a factorization scheme, i.e.,  

$$M^{\mu\nu}(p,q)=Tr[\gamma\cdot pT\gamma\cdot \hat{k} 
HNH^*\gamma\cdot \hat{k} T^*]$$
$$=Tr[T^*\gamma\cdot pT\gamma\cdot \hat{k}]Tr[HNH^*\gamma\cdot \hat{k}], 
\eqno(29)$$
where $\hat{k}$ is on-shell momentum of the propagator $F$.
The first factor in Eq. (29) is the trace of the (soft) forward virtual
quark-target scattering amplitude, and the second factor relates to the  
(hard) probe-parton subprocess.

	On the other hand, the backward
propagator implies the contact (or instantaneous) correlation and we 
can not re-stipulate the backward propagator in any factorization 
scheme. Therefore, one of the factorization conditions is that
we can avoid the backward propagators between 
$T$ and $H$. For this end, we note that a backward parton always 
couples with a pair of forward partons according to Fig. 4.  After 
"pulling down" two forward propagators from the target matrix, the 
backward propagator can be isolated into a hard-subgraph of a next twist 
amplitude. In consequence, we get the twist-3 (Figs. 8b and 8c) and 
twist-4 contributions (Fig. 8d), respectively. In this case, all initial 
partons are forward moving along the direction of target and we can 
break and recombine those parton lines between two factorized parts in 
the factorization scheme. Through the above TOPT-expansion, we also show 
a clear picture where the different twist contributions are separated.
	
\begin{center}
\section{Summary}
\end{center}

In this work, we presented following physical picture about the inverse-power
corrections to the hadron structure in parton model:
(i) a hadron is assumed to be consisted by forward initial partons along 
the target momentum in the infinite momentum frame (i.e., the positively 
of initial parton momentum) and the transverse momentum of parton can be 
neglected if comparing with its longitudinal momentum; (ii) a probe strikes 
with a parton and forms a covariant Feynman propagator connecting with 
the target matrix and probe; (iii) the Feynaman propagator decomposes to 
the forward and backward propagators in TOPT and the backward propagator 
"pulls down" two forward propagators from the target matrix; (iv) after 
factorization, two-parton-correlation function has the dimension of 
$[length]^{-2}$, therefore, the hard-amplitude of 
two-parton-to-two-parton will provide a contribution with $1/Q^2$ to the 
total amplitude. This leads an inverse power correction to the hadron 
structure function.

	In summary, we have generalized time-ordered perturbative theory 
to the collinear limit, where the collinear singularities are eliminated 
in a natural way. We find that a clear space-time picture of the twist-4 
amplitude is presented in its TOPT-form. In particular, we identify the 
special propagators with the backward propagators of partons. Our method
provides a new way to analyze the twist-4 processes in a consistent
QCD-improved parton model.

\noindent {\bf Acknowledgments}:
We would like to thank J.W. Qiu for his encouragement and enlightening 
discussions. We would also like to acknowledge D. Indumathi 
for useful comments. This work was supported by National Natural Science 
Foundation of China and `95-climbing' Plan of China.

\newpage
\noindent {\bf Figure Captions}

\noindent Fig. 1  A twist-2 photon-hadron scattering process of
${\cal{O}}(\alpha_s)$ in TOPT; 
(a) dominant amplitude with the forward propagator and (b) 
suppressed amplitude with the backward propagator.

\noindent Fig. 2  A twist-4 photon-hadron scattering process of
${\cal{O}}(\alpha_s)$ in TOPT; 
(a) suppressed amplitude with the forward propagator and (b) 
dominant amplitude with the backward propagator.

\noindent Fig. 3  Some forbidden vertices of partons in 
twist-4 photon-hadron scattering amplitude. F and B indicate 
forward- and backward-moving partons, respectively.

\noindent Fig. 4  Elemental vertices in the collinear approximation.

\noindent Fig. 5  A complete set of quark-gluon-to-quark-gluon TOPT diagrams,
which give leading power corrections to the hadron structure function.

\noindent Fig. 6  Non-vanished two-quark-to-two-quark TOPT diagrams,
which give leading power corrections to the hadron structure function.

\noindent Fig. 7  Interference processes corresponding to Fig. 5c.

\noindent Fig. 8  TOPT-expansion of the different twist contributions.

\newpage
\begin{thebibliography}{99}

\bibitem{1} A. De Rujula, H. Georgi and H.D. Polotzer, Ann. Phys. (N.Y.)
$\bf 103$, 315 (1977); H.D. Politzer, Nucl. Phys. $\bf B172$, 349 (1980);
S.J. Brodsky, E. L. Berger and G.P. Lepage, in 
Proceedings of the Workshop on Drell-Yan Processes,
Batavia, Illinois, 1982 (Fermilab, Batavia, 1983), p. 137.

\bibitem{2} R.L. Jaffe and M. Soldate, Phys. Lett. 105B (1981) 467;
Phys. Rev. $\bf D26$, 49 (1982).

\bibitem{3} R.K. Ellis, W. Furmanski and R. Petronzio, Nucl. Phys. 
$\bf B207$, 1 (1982); $\bf B212$, 29 (1983).

\bibitem{4} J.W. Qiu, Phys. Rev. $\bf D42$, 30 (1990).

\bibitem{5} G. Altarelli and G. Parisi, Nucl. Phys. $\bf B126$, 298 (1977).
 
\bibitem{6} S.J. Brodsky, Lecture at the 
Summer Institute on Particle Physics at SLAC, California, 1979.

\bibitem{7} X.F. Guo and J.W. Qiu, The leading power corrections to 
the structure functions, BNL/HET-98/36.

\bibitem{8} J.C. Collins, D.E. Soper and G. Sterman, in Perturbative
Quantum Chromodynamics, edited by A.H. Mueller (World Scientific, Singapore)
(1989) 1; J.C. Collins and D.E. Soper, Nucl. Phy. B194 (1982) 445.
    


\end{thebibliography}
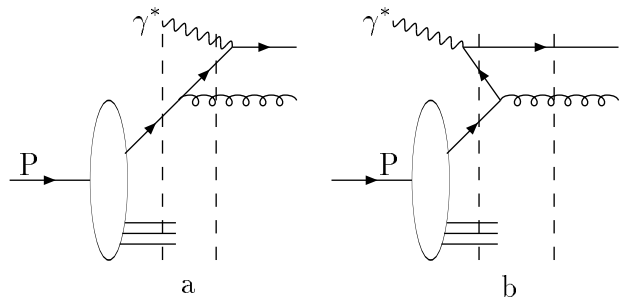
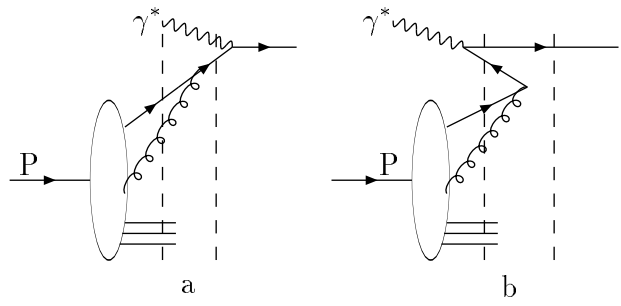
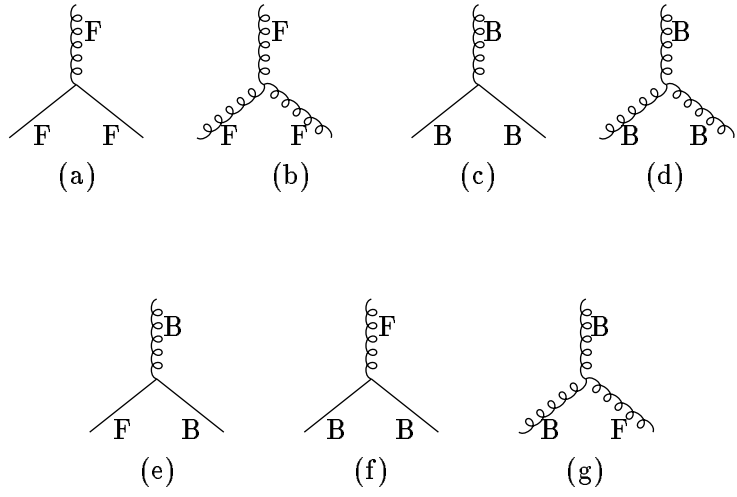
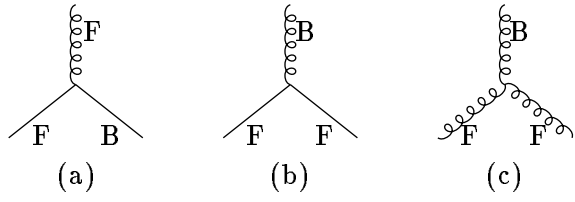
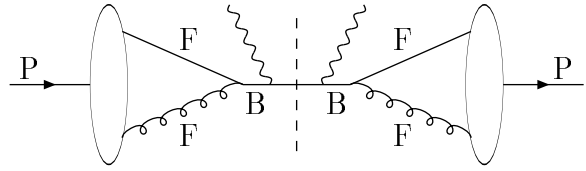
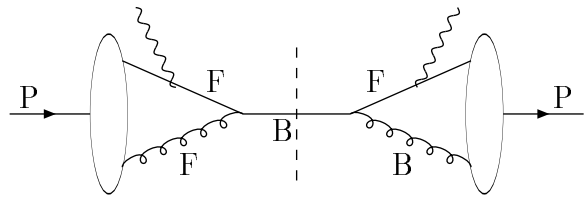
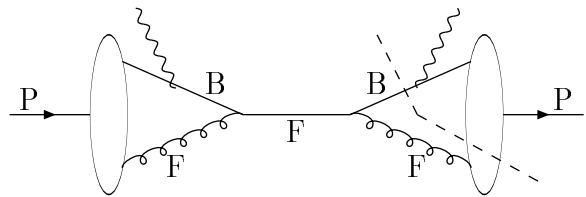
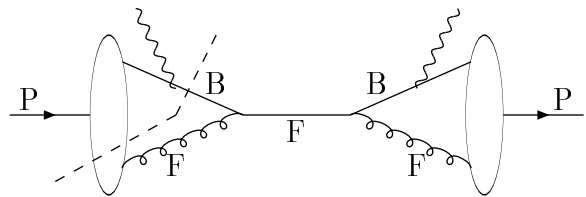
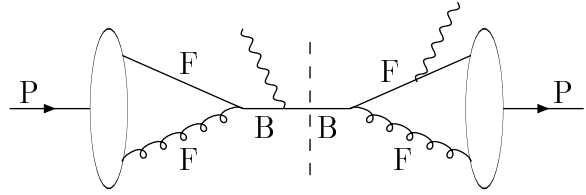
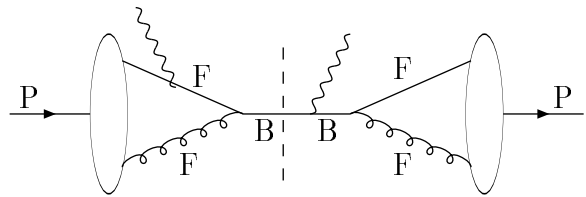
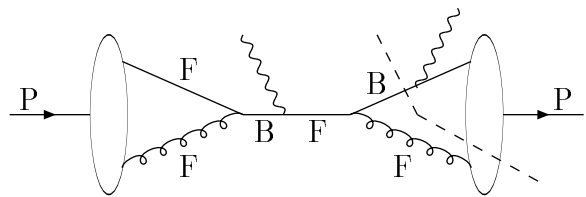
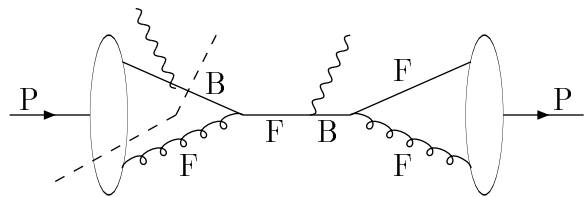
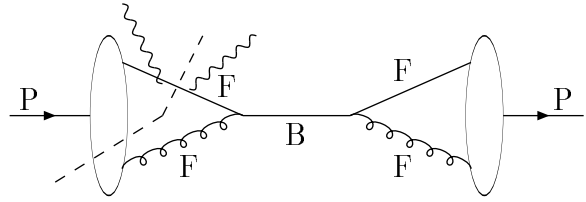
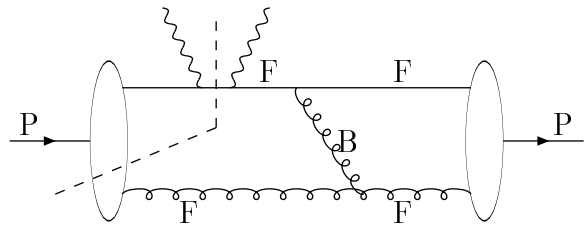
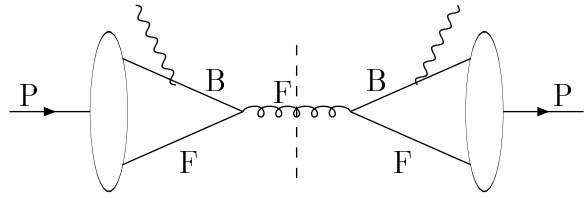
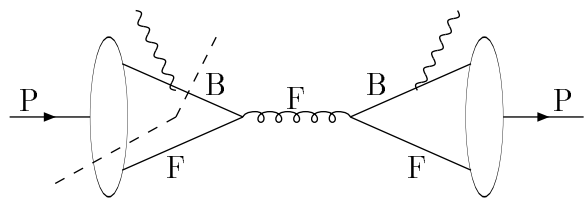
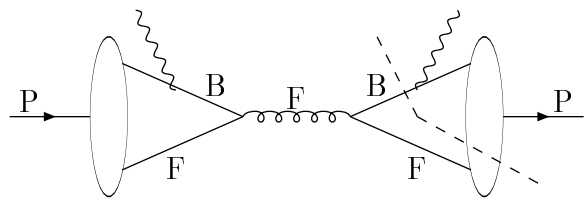
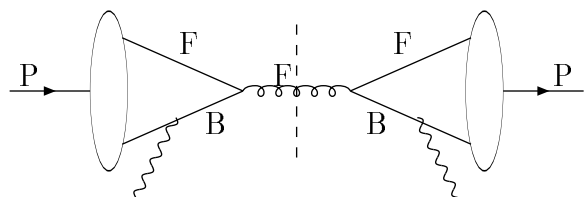
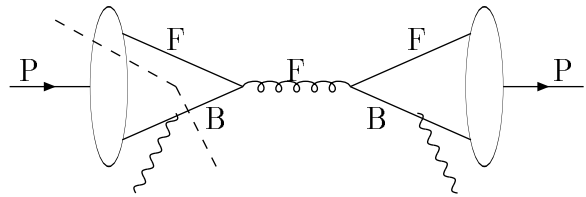
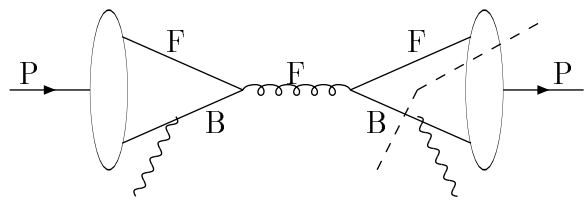
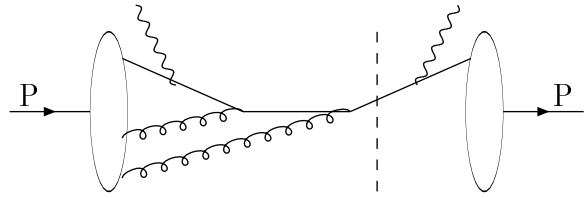
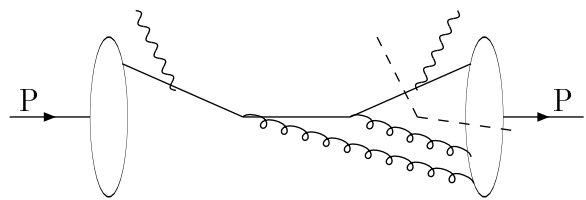
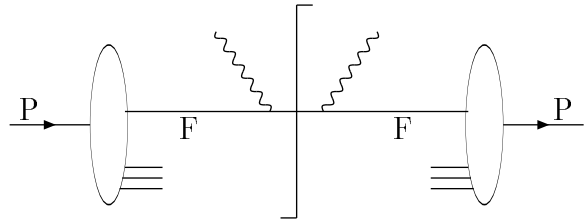
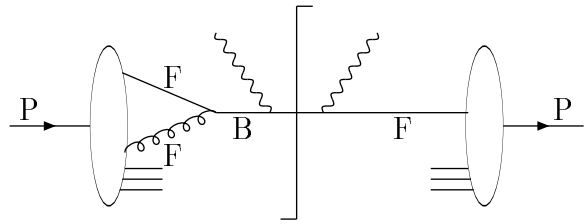
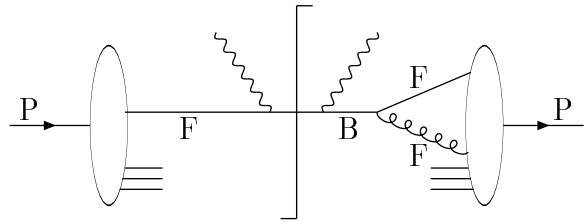
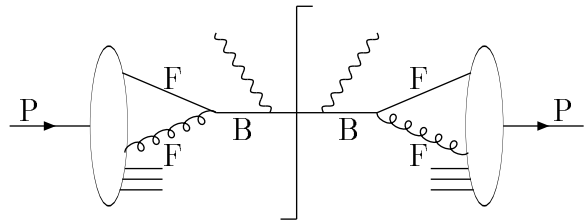
\end{document}